\documentclass[journal]{IEEEtran}
\usepackage{graphicx}
\usepackage{amsmath}
\usepackage{bm}
\usepackage{hyperref}
\usepackage{url}
\usepackage{soul} 
\usepackage{xcolor} 
\usepackage{orcidlink}

\usepackage{cite} 
\begin{document}

\title{A RIS-Enabled Computational Radar Coincidence Imaging}
\author{%
Kavian Zirak\,\orcidlink{0009-0005-6114-618X},~\IEEEmembership{Graduate Student Member,~IEEE}, 
and Mohammadreza F. Imani\,\orcidlink{0000-0003-2619-6358},~\IEEEmembership{Member,~IEEE}%

        
\thanks{K. Zirak and M. F. Imani are with the School of Electrical, Computer, and Energy Engineering, Arizona State University, Tempe, AZ, USA (e-mail: \url{kzirak@asu.edu}; \url{Mohammadreza.Imani@asu.edu}).}}



\maketitle

\begin{abstract}
This paper introduces an innovative imaging method using reconfigurable intelligent surfaces (RISs) by combining radar coincidence imaging (RCI) and computational imaging techniques. In the proposed framework, RISs simultaneously redirect beams toward a desired region of interest (ROI). The interference of these beams forms spatially diverse speckle patterns that carry information about the entire ROI. As a result, this method can take advantage of the benefits of both random patterns and spotlight imaging. Since the speckle pattern is formed by directive beams (instead of random patterns typically used in computational imaging), this approach results in a higher signal-to-noise ratio (SNR) and reduced clutter. In contrast to raster scanning, which requires the number of measurements to be at least equal to the number of unknowns, our proposed approach follows a computational imaging framework and can obtain high-quality images even when only a few measurements are taken. Using numerical simulation, we demonstrate this method's capabilities and contrast it against other conventional techniques.  The proposed imaging approach can be applied to security screening, wireless user tracking, and activity recognition.
\end{abstract}


\IEEEpeerreviewmaketitle

\section{Introduction}

\IEEEPARstart{M}{\lowercase{i}}crowave imaging and sensing can penetrate optically opaque layers or obstacles while preserving users' privacy and comfort \cite{soumekh1994fourier}. As a result, they have been used for security screening at points of interest (e.g. airports), to detect users' presence or biomarkers, or recognize their gestures \cite{sheen2001three,li2019intelligent,nanzer2017review}. The primary issue in conventional microwave imaging and sensing systems is the need to realize a large aperture to achieve the desired resolution. This problem can be overcome by using mechanically scanned antennas, large antenna arrays, or a combination of both. While delivering satisfactory results, these techniques usually result in costly, slow, and bulky systems \cite{sheen2001three,ahmed2021microwave,moulder2016development,daum2008radar}.

To overcome this issue, recent works have combined compressive sensing \cite{donoho2006compressed} and metamaterial antennas \cite{gollub2017large,imani2020review,venkatesh2020high} to reduce the cost and complexity of close-range microwave imaging systems. In these works, metasurface antennas are designed to generate spatially diverse patterns as a function of frequency or electronic signals that encode the reflectivity map of the ROI into a few measurements. These measurements were then computationally processed to deduce the reflectivity map. In this manner, a few metasurface antennas could replace large antenna arrays or mechanical scanning.

Recently, reconfigurable intelligent surfaces (RISs) have gained significant attention for their ability to redirect signals and engineer wireless propagation \cite{renzo2019smart,basar2019wireless,wang2023applications,elmossallamy2020reconfigurable,trichopoulos2022design,bae2024overview,zhang2024secure,alamzadeh2021reconfigurable,pei2021ris,gholami2025wireless, gholami2024realization}. Given their electrically large size and reconfigurability, they also hold great potential for realizing a simple and cost-efficient microwave imaging system. RIS-enabled imaging offers unique benefits: it allows for using a single transmitter and receiver (close to each other), ensuring simple temporal and phase synchronization, while eliminating the need for switch arrays or mechanical movement, further reducing system complexity and cost.

In previous works, RISs have played two different roles as part of the imaging system: 1) They have been used to generate random spatially diverse patterns \cite{wang2016single,tao2020distributed,jiang2024near,li2019intelligent,wang2022intelligent,wang2025dreamer}. This configuration closely resembles the metasurface-enabled computational imaging configurations \cite{gollub2017large,imani2020review}. As a result, this method offers similar advantages. For instance, the RISs do not need to generate directive beams, simplifying the design process. By utilizing computational techniques, a large imaging area can be reconstructed using just a few measurements, thereby increasing the data acquisition rate.
However, random patterns are not directive, which can result in a lower SNR. This can be especially problematic given the two-way propagation loss when using RISs \cite{zhang2022active,wang2025dreamer}. They are also susceptible to clutter; random patterns can be scattered by objects not in the imaging domain, degrading performance. 2) Alternatively, RISs can form directive beams to scan an imaging domain. In this manner, the system can enjoy a large SNR due to the directivity of the large RIS. It also reduces the impact of clutter by directing the beam toward the desired imaging domain. However, raster scanning a directive beam over the ROI can be time-consuming. 

\begin{figure}[t]
    \centering
    \includegraphics[width=0.45\textwidth]{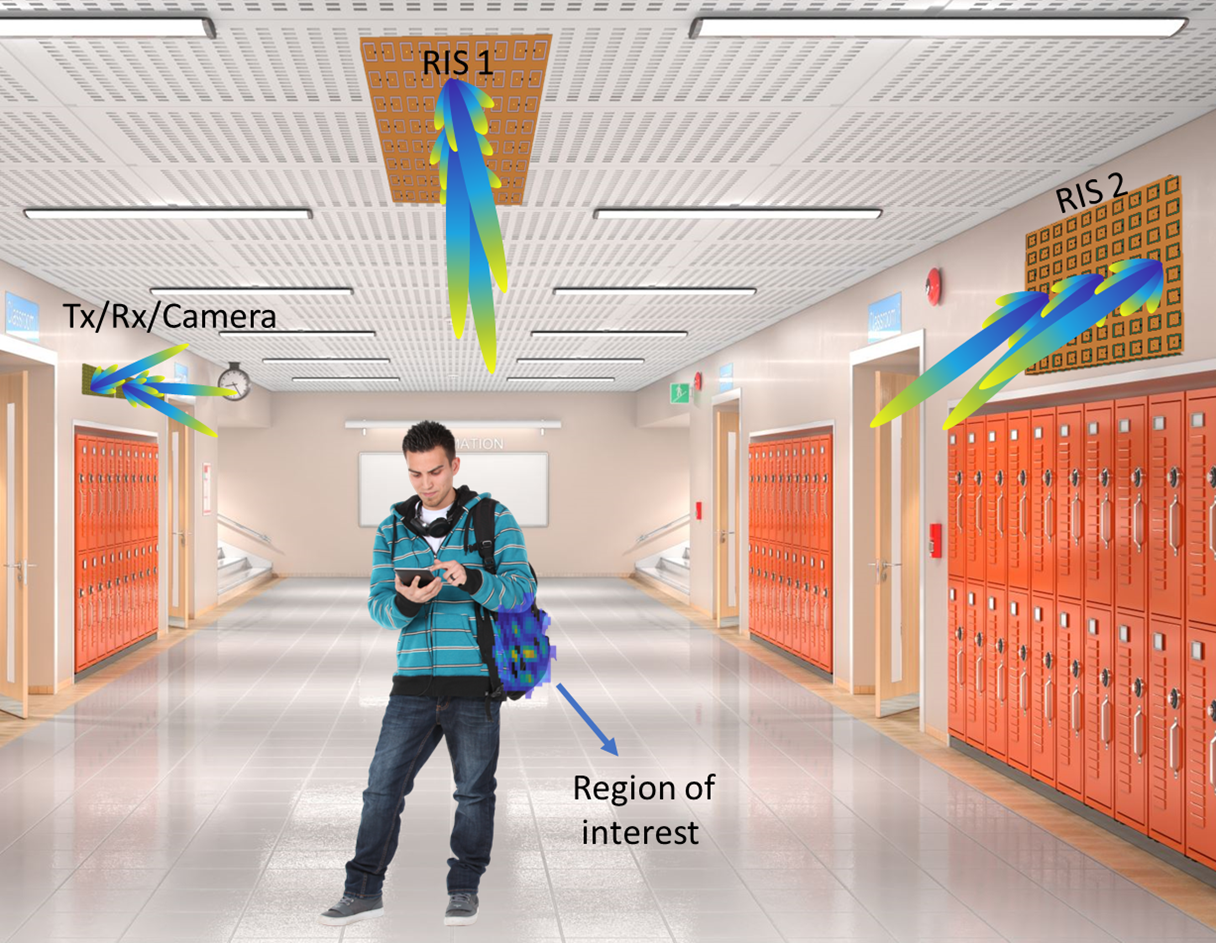}
    \centering
    \caption{Proposed RIS-Enabled microwave imaging approach.}
    \label{fig:f0}
\end{figure}

In this work, inspired by radar coincidence imaging (RCI) \cite{diebold2019phaseless}, our objective is to combine the benefits of computational imaging using random patterns with the high SNR and low clutter of directive beam raster scanning \cite{boyarsky2017synthetic}.  In our method, we assume a transmitter (Tx) emits electromagnetic waves toward a collection of RISs (or different portions of a large RIS), which redirect it toward the ROI, as visualized in Fig. \ref{fig:f0}.  This ROI is either predefined or has been identified by an optical or IR camera. The interference of these redirected beams over the ROI produces speckle patterns that encode spatial information across the ROI. We can change the resulting speckle pattern by modifying the RISs' response.  The reflected waves from the object are then redirected by another set of RISs toward the receiver (Rx) antenna. The \textit{focused yet spatially diverse patterns} of the transmitting and receiving RISs can encode the information of the ROI, which can be processed computationally to reconstruct the ROI’s reflectivity. 

In \cite{zirak2025novel}, we presented preliminary results supporting this concept. Building upon that work, we will detail the proposed imaging approach configuration and illustrate the process to generate \textit{focused speckle patterns}. The forward and inverse models used to analyze the imaging performance will be outlined. Using this model, we contrast images obtained by the proposed method against those from raster scanning and random patterns. We conclude the paper with a summary of findings and a discussion of future research directions.

\section{Proposed Imaging Method} \label{sec:method}
\begin{figure}[ht]
    \centering
    \includegraphics[width=0.45\textwidth]{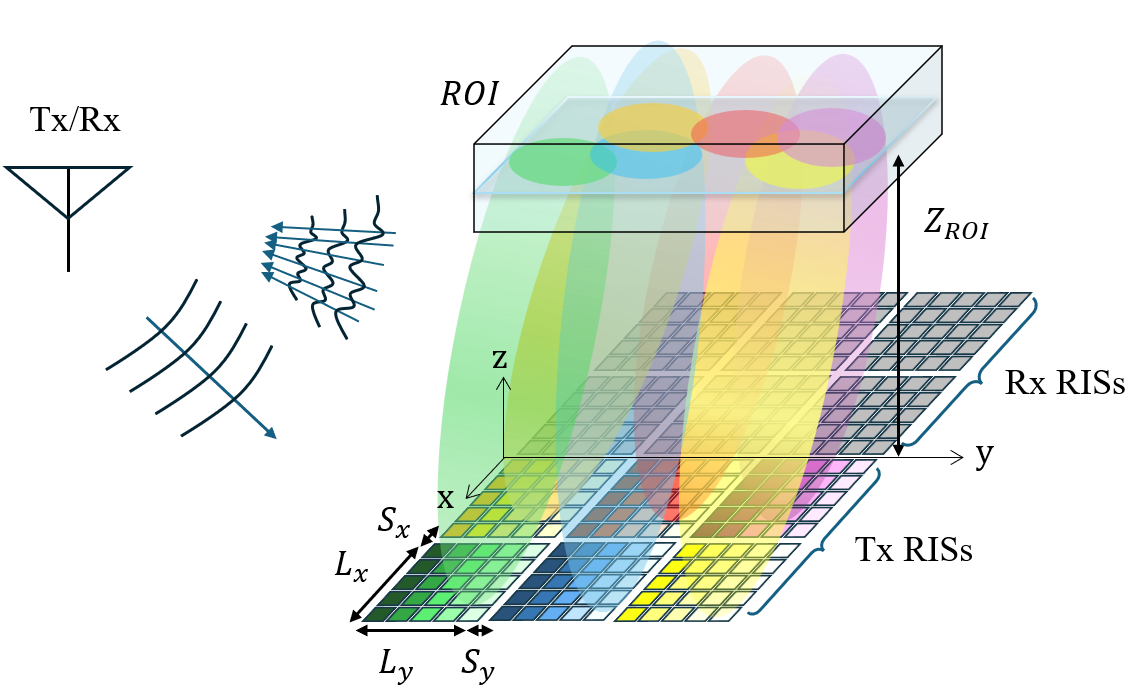}
    \centering
    \caption{The general configuration of the imaging problem.}
    \label{fig:f1}
\end{figure}

To demonstrate the proposed method numerically, we use the configuration shown in Fig. \ref{fig:f1}. This configuration consists of six RISs (Tx-RISs) that redirect signals from the transmitter and six RISs (Rx RISs) that redirect the scattered signals from ROI toward a receiver (Rx). For this demonstration, we assume each RIS consists of a an array of $20\times20$ elements, with a resonance frequency of 6 GHz and an element spacing of 2 cm ($\lambda/2.5$), resulting in an overall size of $L_x = L_y = 40 \, \text{cm}$. Each RIS element is assumed to be loaded by a varactor diode for continuous phase adjustment of the reflected signal. For this demonstration, we assume that the varactor diode can change the phase of the reflected signal over its full range and that it exhibits no losses. In this manner, we can compare the performance of different imaging schemes without limitations caused by specific elements or varactor choices. To simplify the calculations, we also assume no coupling between the elements. The distance between the edges of RISs along the $x$ and $y$ axes is represented by $S_x=0$ and $S_y=0$ , respectively. The transmitter and the receiver antennas are assumed to be collocated ($x = 0$, $y = -3 \, \text{m}$, $z = 3 \, \text{m}$).

\begin{figure}[t]
    \centering
    \includegraphics[width=0.48\textwidth]{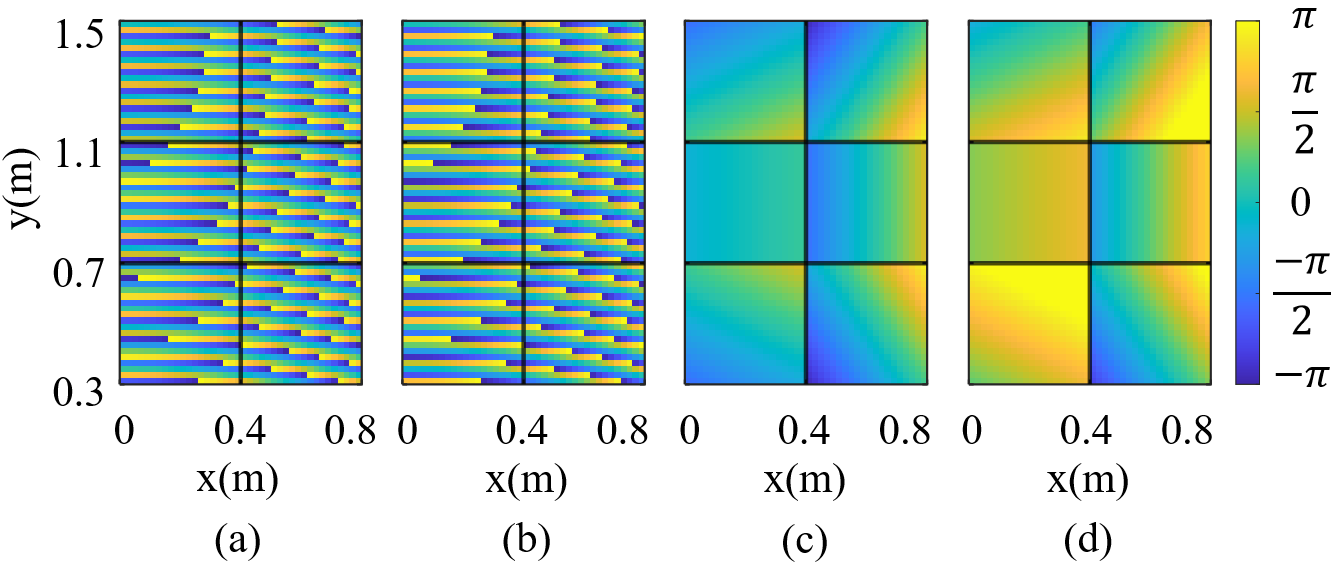}
    \centering
    \vspace{-0mm}
    \caption{(a) Phase accumulation from Tx antenna to the RIS plane. (b) Phase compensator. (c) Phase gradient. (d) The same phase gradient as in (c) but with different phase offsets.}
    \label{fig:f2}
\end{figure}

The transmitter emits a spherical wave toward the Tx-RISs. Fig. \ref{fig:f2}a shows the phase accumulated by this signal as it arrives on each RIS's elements. This profile needs to be compensated (by RIS's phase responses) to redirect this signal toward arbitrary directions, as shown in Fig. \ref{fig:f2}b. We then add a gradient phase term to the RIS elements to redirect the beams in a specific direction (defined as $\theta$ and $\phi$). The phase gradient term is given by:

\begin{equation}
\phi_{d,mn} = k_0 \left( x_{mn} \sin \theta \cos \phi + y_{mn} \sin \theta \sin \phi \right)
\end{equation}

where $k_0$ represents the propagation constant in free space, and ($x_{mn}$,  $y_{mn}$) are the coordinates of the $mn^{\text{th}}$ RIS element. Fig. \ref{fig:f2}c shows an example of the aggregate phase delay profile, assuming the beams are directed toward a point in the ROI ($x = 0$, $y = 0.9 \, \text{m}$, $z = 8 \, \text{m}$). 

The underlying idea in our proposed scheme is to obtain information from each Tx-RIS and Rx-RIS pair while they are used simultaneously. Toward this goal, we use all RISs to direct signals toward the ROI but with different phases and directions. In this manner, we can form a speckle-looking pattern confined to the ROI. To enable that, we use two design knobs: 1) We add a random phase offset to all the elements of each RIS. This process is visualized in Figs. \ref{fig:f2}c and d. Both phase gradients are designed to redirect the beam toward the same directions, but with different phase offsets. 2) We introduce a random component to the desired redirection angles ($\theta,\phi$). These random directions are defined such that the redirected beam is still within the ROI. The random component for each redirection angle is calculated using:
\begin{equation}
    \theta_{\text{rnd}}
 = C_\theta (\theta_{\text{max}} - \theta_{\text{min}}),\;\;\mathrm{ 
 }\phi_{\text{rnd}}
 = C_\phi (\phi_{\text{max}} - \phi_{\text{min}}).
    \tag{2}
\end{equation}

\begin{figure}[t]
    \centering
    \includegraphics[width=1\linewidth]{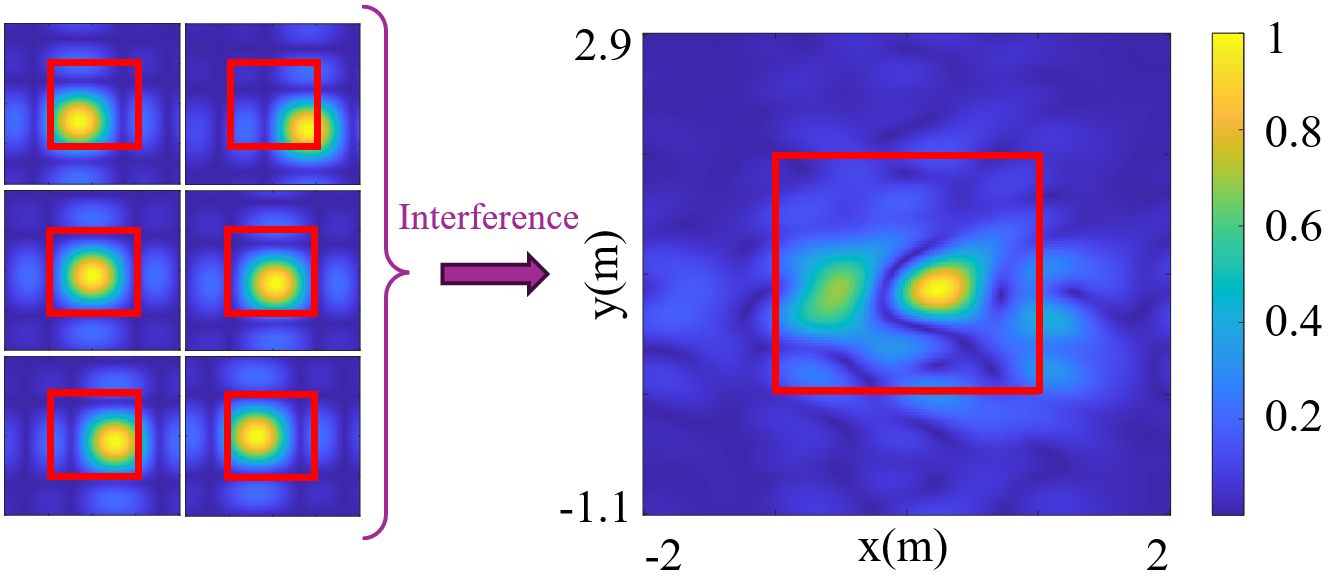}
    \centering
    \caption{Normalized field patterns formed by individual RISs (left) and their interference pattern (right).  Red rectangles mark ROI.}
    \label{fig:f2_2}
\end{figure}

Here, the terms \(\theta_{\text{max}}\) and \(\theta_{\text{min}}\) are the maximum and minimum elevation angles calculated such that the redirected beam stays within the ROI for a certain RIS. The same goes for \(\phi_{\text{max}}\) and \(\phi_{\text{min}}\). \(C_\theta\) and \(C_\phi\) are random values drawn between 0 and 0.25--- ensuring the beam remains within the ROI boundaries. Fig. \ref{fig:f2_2} visualizes the proposed method to form a speckle pattern confined to the ROI. Specifically, it shows the field amplitude distribution in the ROI (denoted by a red rectangle) for each of the six Tx-RISs and their interference. Due to different directions, phase offset, and location, the interference of these beams results in a speckle pattern.

\begin{figure}[t]
    \centering
    \includegraphics[width=1\linewidth]{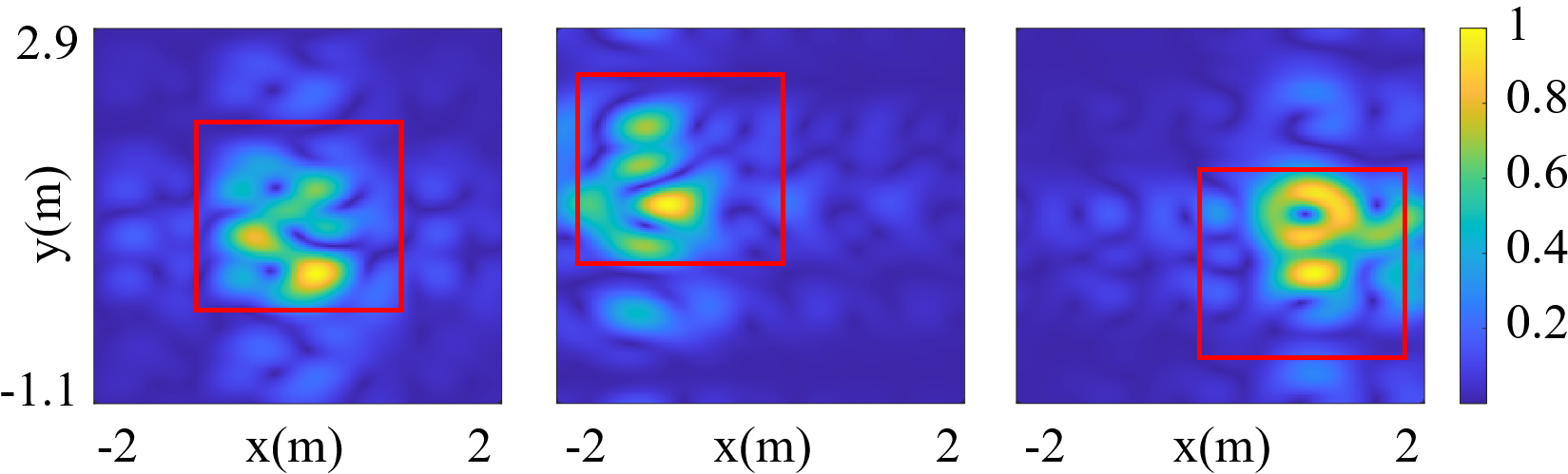}
    \centering
    \caption{Demonstration of the method's adaptability across three ROIs.}
    \label{fig:f4}
\end{figure}

Fig. \ref{fig:f4} demonstrates the possibility of adjusting the focused speckle pattern to different ROIs. In practice, one can optimize the value of the scaling factor ($C_\theta$ and $C_\phi$) for different directions of the ROIs, or fine-tune it empirically to ensure that the beams remain confined within the desired region. 

\begin{figure}[t]
    \centering
    \includegraphics[width=\linewidth]{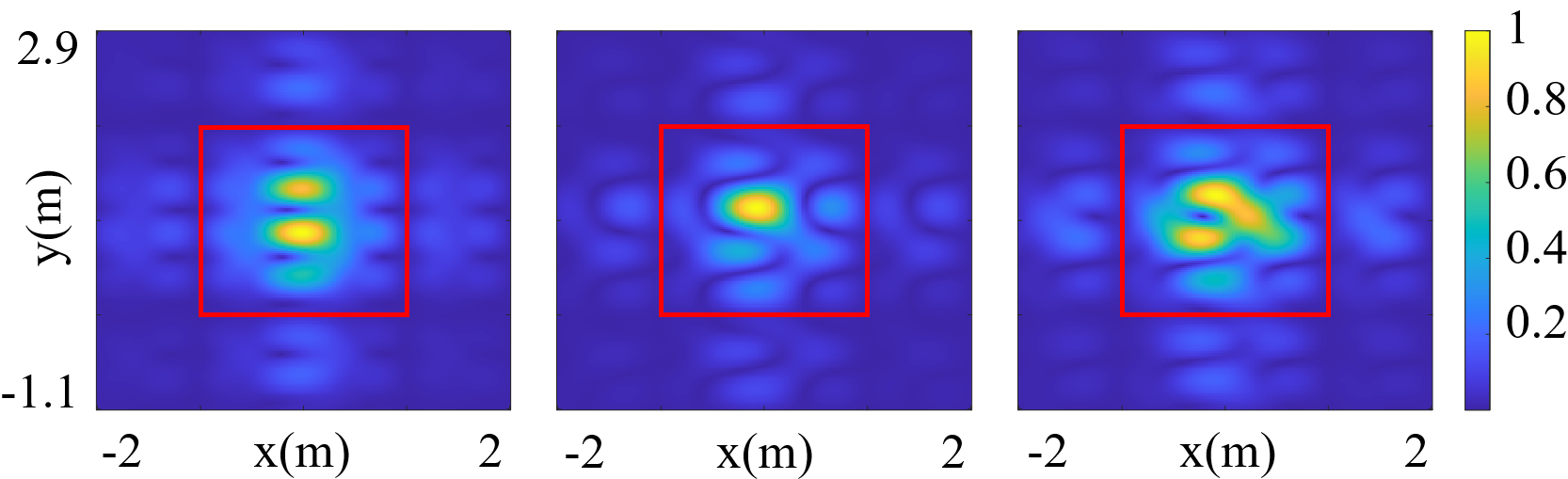}
    \centering
    \caption{Speckle patterns generated by varying the offset while keeping the frequency and redirection angle fixed.}
    \label{fig:f5_offset}
\end{figure}

\begin{figure}[t]
    \centering
    \includegraphics[width=0.87\linewidth]{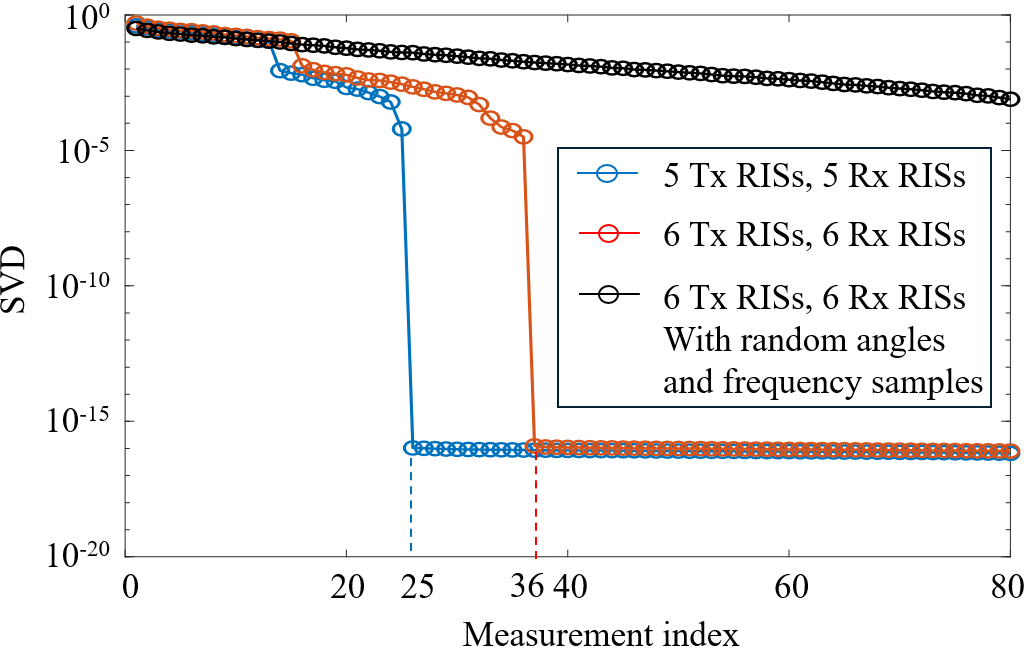}
    \centering
    \caption{Comparison of normalized SVD curves across different configurations.}
    \label{fig:f6_svd}
\end{figure}

\begin{figure}[t]
    \centering
    \includegraphics[width=0.95\linewidth]{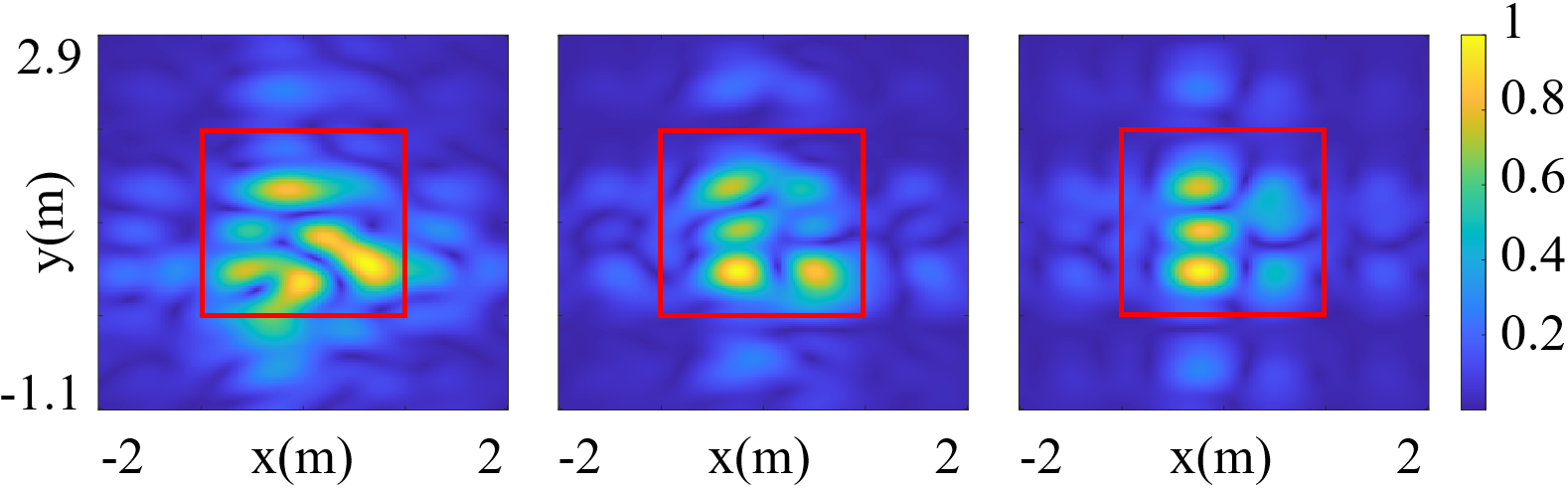}
    \centering
    \caption{Speckle patterns generated by varying the redirection angle while keeping the offset and frequency fixed.}
    \label{fig:f8_angles}
\end{figure}

For brevity, we will refer to each set of redirection angles and phase offsets across the RISs as a \textit{mask}. The Tx-RISs produce a unique, focused speckle pattern for each mask and frequency. To illustrate the impact of each parameter individually, Fig. \ref{fig:f5_offset} shows the speckle patterns when only the offset varies, with the frequency and redirection angle fixed. Clearly, as the phase offset of the Tx-RIS changes, the resulting speckle pattern also changes. However, the number of unique patterns that can be generated by using random phase offsets is constrained by the product of the number of Tx and Rx RISs. As shown in Fig. \ref{fig:f6_svd}, the singular value decomposition (SVD) of the resulting patterns drops dramatically once the number of measurements exceeds this limit. To overcome this issue, we also randomly change the redirection angles. Fig. \ref{fig:f8_angles} illustrates this case: While the beam redirected from each RIS is randomly shifted within the ROI, it leads to significant variation in the spatially diverse patterns still confined to the ROI. This is also reflected in the SVD curves of Fig. \ref{fig:f6_svd}.

\begin{figure}[t]
    \centering
    \includegraphics[width=0.95\linewidth]{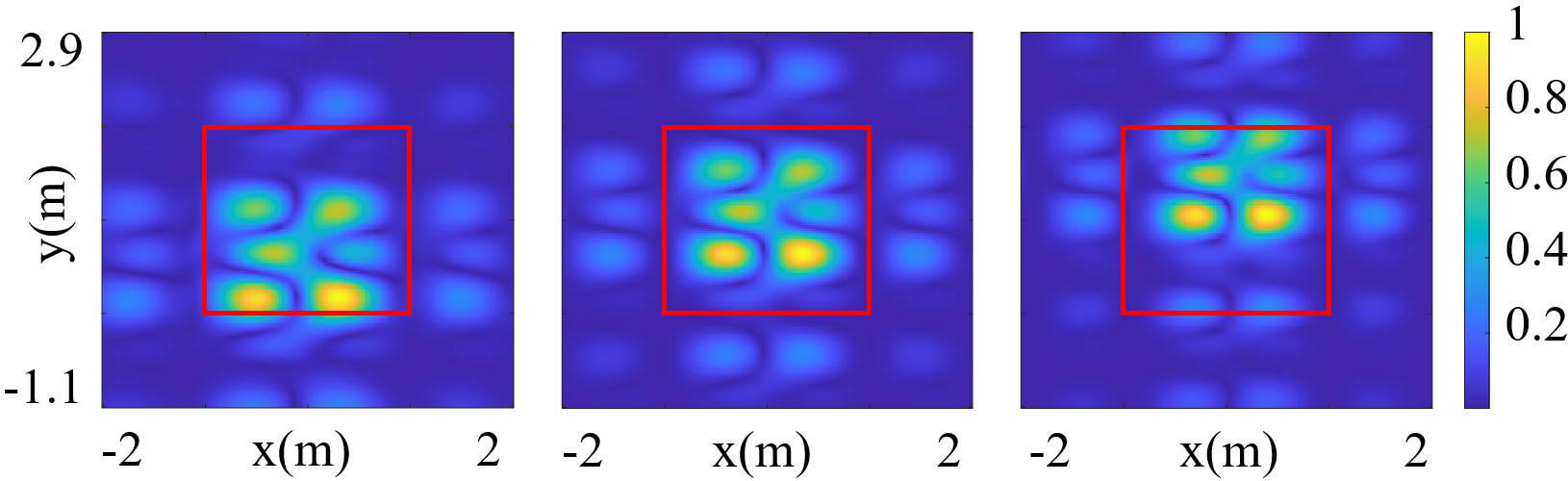}
    \centering
    \caption{Speckle patterns generated by the Tx-RISs as the frequency is changed to 5.9 GHz, 6 GHz, and 6.1 GHz from left to right.}
    \label{fig:f7_freq}
\end{figure}

The change in frequency plays two key roles: 1) Using a bandwidth allows for detection in the range direction. 2) Changing frequency results in frequency diverse patterns, which can also be interpreted as filling up the k-vector space to reduce aliasing, as discussed in \cite{marks2016spatially}. The impact of frequency on the patterns is visualized in Fig. \ref{fig:f7_freq}. It illustrates the squinting of the patterns as the frequency changes---while the phase offset and redirection angle are kept constant. 

To demonstrate the utility of focused speckle patterns in forming images, we have developed a simplified imaging simulator based on \cite{lipworth2015comprehensive}. In the forward model, we assume the target's location is known and discretized densely into $P$ pixels (for a 2D target), represented by the reflectivity map \( \mathbf{\sigma}_{P \times 1} \). Assuming the First-Born approximation, the relationship between the reflectivity of the target and the measured signal, $\mathbf{g}_{M \times 1}$, is given by:

\begin{equation}
    \mathbf{g}_{M \times 1} = \mathbf{H}^{\text{FWD}}_{M \times P} \boldsymbol{\sigma}_{P \times 1},
    \label{eq:reflectivity_map}
    \tag{4}
\end{equation}
where $M$ is the number of measurements, determined by the product of frequency points and the number of masks. The forward sensing matrix \( \mathbf{H}^{\text{FWD}} \) is computed by evaluating the fields generated by the RISs at each pixel location:

\begin{equation}
    h_{\text{fwd}}(i, j) = \vec{E}_{\text{Tx}}^i (\vec{r}_j) \cdot \vec{\vec{E}}_{\text{Rx}}^i (\vec{r}_j),
    \label{eq5}
\end{equation}
where $i$ represents the $i$-th measurement and \( \vec{r}_j \) is the location of the $j$-th pixel. In the inverse model, we discretize the ROI around the target into $N$ voxels and computed the inverse sensing matrix, $\mathbf{H}^{\text{INV}}_{M \times N}$, for the same $M$ measurement configurations, in a similar manner as in equation (\ref{eq5}). Using the inverse sensing matrix, $\mathbf{H}^{\text{INV}}_{M \times N}$ and $\mathbf{g}_{M \times 1}$, we solve for the reflectivity of the target, $\boldsymbol{\sigma}_{N \times 1}$ of the whole ROI. Since the inverse sensing matrix $\mathbf{H}^{\text{INV}}_{M \times N}$ is not square and usually has few number of measurements than the unknown reflectivity ($M<N$), we use computational techniques (e.g., CGS), to obtain the estimated reflectivity map \( \Tilde{\mathbf{\sigma}}_{N \times 1} \). 

\begin{figure}[t]
    \centering
    \includegraphics[width=\linewidth]{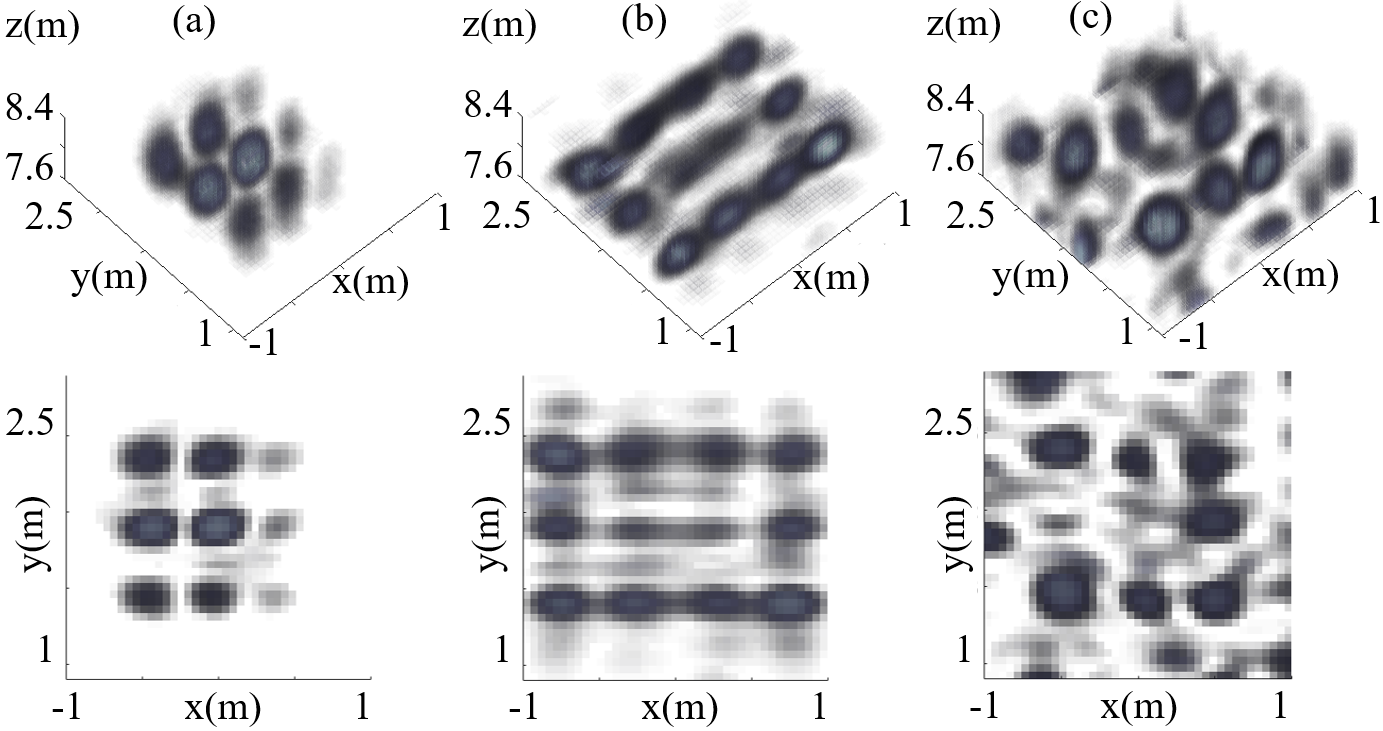}
    \centering
    \caption{Reconstructed images using (a) our method, (b) raster scanning, and (c) random patterns in 3D (top row) and front view (bottom row).}
    \label{fig:f9_comp3}
\end{figure}

Figure \ref{fig:f9_comp3} compares the proposed imaging method with raster scanning and random pattern techniques, all using five frequency points ranging from 5.9 GHz to 6.1 GHz and 20 masks or scans, resulting in a total number of 100 measurements. To implement random patterns, we assign a random phase to each element of the RIS. For raster scanning, we used all 6 Tx RISs (or 6 RX RISs) to direct the beam toward a given location inside the RIS and then raster scan the targeted voxel within the ROI. For this demonstration, we consider the targets located at $z = 8 \, \text{m}$. The target consists of 9 subwavelength scatterers arranged in a $3 \times 3$ grid, with a spacing of $50~\mathrm{cm}$ along both the $x$- and $y$-axes, centered around $x = 0$ and $y = 190~\mathrm{cm}$. We also assume $20$dB SNR. In the raster scanning case, the number of measurements is limited, which leads to poor reconstruction quality. The resulting image appears stretched and aliased, with the system failing to resolve the correct number and shape of targets, suggesting undersampling and insufficient spatial frequency coverage. In the random pattern approach, although more spatial diversity is present, the lack of beam directivity results in a low SNR, producing a highly noisy reconstruction and reduced target visibility. In contrast, our proposed method achieves significantly better reconstruction quality, accurately localizing the targets within the region of interest while avoiding artifacts caused by aliasing or noise. 

\begin{figure}[t]
    \centering
    \includegraphics[width=\linewidth]{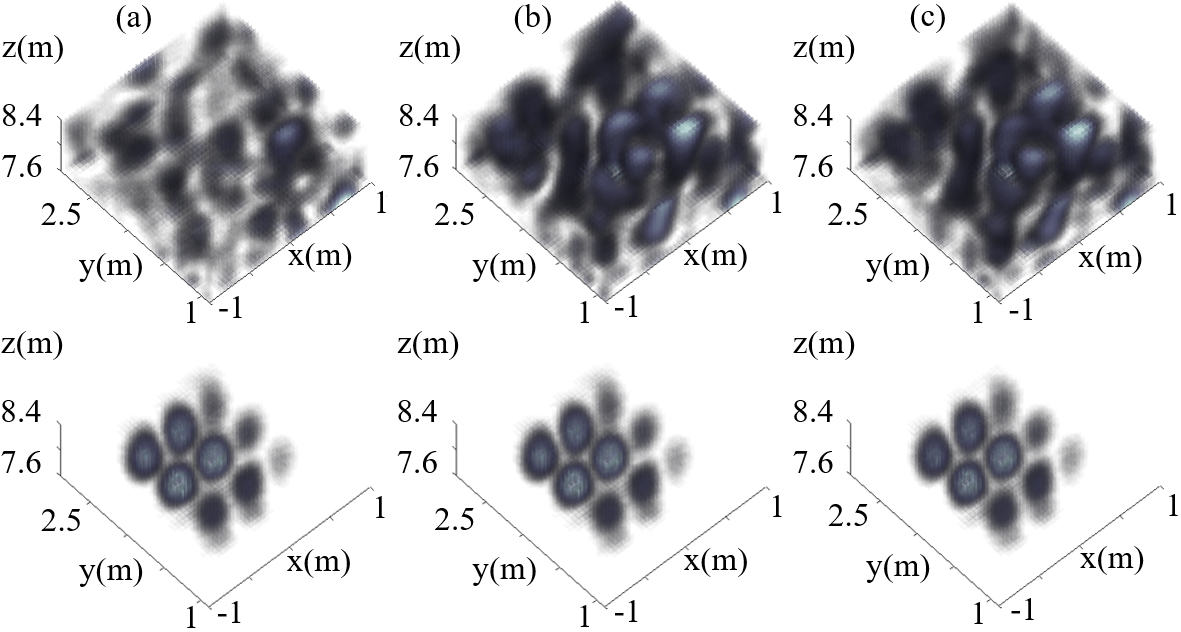}
    \centering
    \caption{Reconstructed images using random patterns (top row) and our proposed method (bottom row) in the presence of a clutter object located at $x=0, y = 3, 5, 10, z=8$ m (left to right).}
    \label{fig:f11_clutter}
\end{figure}

To further demonstrate the advantages of our proposed method, we compare its robustness to clutter against the random pattern approach. Figure \ref{fig:f11_clutter} shows the reconstruction results in the presence of a clutter object located at $y = 3, 5 , \mathrm{or} 10$ m, using 30 masks and an SNR of 50 dB. The top row presents results from the random pattern method, where strong degradation is observed when the clutter is near the region of interest. While some improvement is seen as the clutter moves farther away, the reconstruction remains highly distorted. In contrast, the bottom row shows reconstructions using our proposed method, which remain robust to clutter. These results highlight the method’s inherent resilience to out-of-region scattering and its ability to suppress clutter effects. Such a feature becomes useful when using RISs to image a small ROI in a crowded region as envisioned in Fig. \ref{fig:f0}.

While we have illustrated the performance of the proposed method using an ideal model for RISs, the setting has been kept the same between different methods to present a fair comparison. In practice, phase quantization and other losses can result in beam quality degradation. However, both issues would degrade all methods. It is worth noting that the presented method can be adapted for targets at different distances. The only reason to use 8m distance was to ensure the target was far away from the aggregate 6 RIS so that a directive beam is formed for raster scanning. Our proposed method can easily be used for smaller ranges. The proposed method does not need separate RISs; a single large RIS can be divided into virtual distinct RISs. In fact, changing the size of the virtual panels can also be used to increase diversity.

\section{Conclusion} \label{sec:conclusion}
We proposed a novel method for RIS-enabled imaging by combining spotlight with compressive imaging techniques. Using a numerical model, we demonstrated that our proposed method benefits from the strengths of spotlight imaging, particularly its robustness in cluttered environments. We also showed that, similar to other computational imaging techniques, our approach can produce high-quality images using a few measurements. This proposed method paves the way for utilizing RISs for imaging and sensing in crowded environments.

\section*{Acknowledgment}

This material is based upon work supported by the National Science Foundation under Grant No. ECCS-2333023.

\bibliographystyle{ieeetr}
\bibliography{Bibliography}
\end{document}